# Matter creation via gamma-gamma collider driving by 10 PW laser pulses

Jinqing Yu[1], Haiyang Lu[1,2], T. Takahashi[3], Ronghao Hu[1], Zheng Gong[1], Wenjun Ma[1], Yongsheng Huang[4,5], and Xueqing Yan[1,2,6]

**The nature of matter creation is one of the most basic processes in the universe. According to the quantum electrodynamics theory, matters can be created from pure light through the Breit Wheeler (BW) process[1]. The multi-photon BW process ($\gamma+n\gamma' \rightarrow e^++e^-$) has been demonstrated in 1997 at the SLAC[2], yet the two-photon BW process ($\gamma+\gamma' \rightarrow e^++e^-$) has never been observed in the laboratory. Interest has been aroused to investigate this process with lasers[3,4] due to the developments of the laser technology[5] and the laser based electron accelerators[6]. The laser based proposals[3,4] may be achieved with NIF[7] and ELI[8], provided that the signal-to-noise (S/N) ratio of BW is high enough for observation. Here, we present a clean channel to observe the matter creation via a gamma-gamma collider by using the collimated γ-ray pulses generated in the interaction between 10-PW lasers and narrow tubes. More than $3.2\times10^8$ positrons with a divergence angle of ~7 degrees can be created in a single pulse, and the S/N is higher than 2000. This scheme, which provides the first realization of gamma-gamma collider in the laboratory, would pave the developments of quantum electrodynamics, high-energy physics and laboratory astrophysics.**

## Introduction

The two-photon interaction, which is one of the most fundamental process in the universe and the quantum electrodynamics theory, has not been observed in the laboratory. To realize so, either the ultra-clean interaction environment or the high S/N ratio could enhance the credibility of the experiments. The conventional accelerator based γ-γ collider[2,9] may provide a clean interaction environment due to its controllable, reliable and reproducible of the electron bunches. However, a large machine is needed and the pair production is less 0.1 per shot[9].

---

[1]Key Laboratory of HEDP of the Ministry of Education, CAPT, and State Key Laboratory of Nuclear Physics and Technology, Peking University, Beijing 100871, China. [2]Collaborative Innovation Center of Extreme Optics, Shanxi University, Taiyuan, Shanxi, 030006, China. AdSM Hiroshima University, 1-3-1 Kagamiyama, Higashi Hiroshima, Hiroshima 739-8530, Japan. [4]Institute of High Energy Physics, Chinese Academy of Sciences, Beijing 100049, China. [5]State Key Laboratory of Particle Detection and Electronics (Institute of High Energy Physics, CAS), Beijing 100049, China. [6]Shenzhen Research Institute of Peking University, Shenzhen 518055, China. Correspondence and requests for materials should be addressed to H. Y. L. (hylu@pku.edu.cn) and X. Q. Y. (x.yan@pku.edu.cn).

The laser based γ-γ collider takes the advantage of high efficiency since $10^2 \sim 10^5$ pairs could be created in one shot. O. Pike, et al[3] proposed the first scheme to achieve γ-γ collider in vacuum by the means of the interaction between the energetic γ-ray beam and the radiation field. However, due to the intrinsic divergence of the laser-driven electron beams, a large number of γ-ray photons spread off the focal spot and interact with the hohlraum of high-Z material. In such condition, the noise from the Bethe-Heitler[10] process may overwhelm the signal. In the proposal of X. Ribeyre, et al[4], although the possibility to detect the BW pairs in the preferential direction has been discussed, it is a huge challenge to distinguish the signal ($10^3 \sim 10^5$) from the background positrons ($> 10^{10}$) generated from the processes of multi-photon BW[11], Trident and Bethe-Heitler[10].

In the two-photon BW process, the pair production is a function of the photon number, the collision area size and the cross-section of BW. The laser intensity above $10^{23}$ W/cm$^2$ would be accessible by the under-construction laser facilities[8,12]. In the laser matter interaction under such intensity, a large number ($>10^{13}$) of γ-rays ($>0.511$ MeV)[13,14] can be generated. Such photons could be a perfect source for γ-γ collider if the divergence of the γ-rays and the background positrons can be well reduced. Here, we employ the collimated γ-ray pulses generated from the interaction between narrow tube targets and 10 PW lasers into γ-γ collider. Numerical calculations show that more than $3.2 \times 10^8$ positrons with a divergence of 7° can be generated in one shot. All the possible background positrons generated from the multi-photon BW[11,15], Trident and Bethe-Heitler[10] together with Triplet[16] processes are considered in detail. Results indicate that the BW signal is more than three orders of magnitudes higher than the background level. The signal of collimated positrons is detectable since it is much higher than the threshold of the existing spectrometers[17,18].

Figure 1 shows the setup of γ-γ collider, in which the γ-ray pulses are generated by irradiating the 10 PW lasers into the narrow tube targets. The γ-ray pulses collide outside the target where the electron-positron pairs can be created. The spectrometers, which have been successfully used to detect the laser based positron sources[17,18], can be placed off the collision area for a distance of several cm.

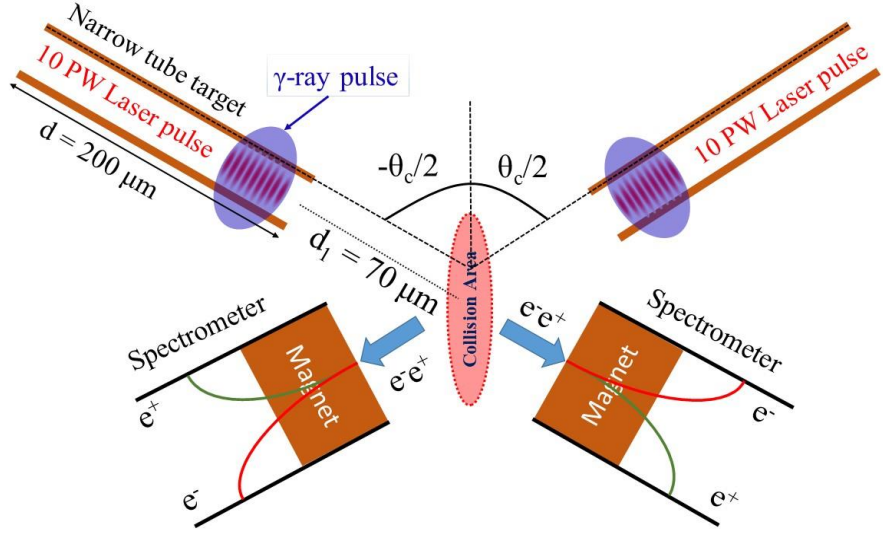

**Figure 1 | Schematic of gamma-gamma collider.** The 10 PW lasers irradiate into the narrow tube targets for the radiation of γ-ray pulses. The γ-ray pulses collide in an angle of $θ_c$. A distance $d_1$= 70μm from the target end to the collision point is considered to avoid the e⁻e⁺ pairs produced from the multi-photon BW process[2,11,19] between the γ-ray photons and the remaining laser pulse from the other side, since the laser intensity greatly decreases after transmitting into the vacuum for a distance of $d_1$. The length of the targets, d, is fixed to 200 µm because most of the photons are generated in 70-200 µm. The effective distance from the γ-ray source original location to the collision point approximates the target length d. The Spectrometers which can separate the electrons and positrons are placed in the directions of the γ-ray pulses.

In the γ-γ collider, more pairs can be generated by the photon beams of higher flux. The rapid development of high power laser technology[5] is paving the way to generate compact, energetic γ-ray sources of high brilliance[20]. Recently, we have demonstrated the generation of a high flux γ-ray pulse through the interaction between an ultra-intense laser pulse of 10 PW and a narrow tube target[21].

With the available parameters of the up-coming lasers[8,12], we simulate the generation of collimated γ-ray pulses by using two-dimensional particle-in-cell code EPOCH2D[22] (for more detail on the simulation setup see methods). Here, we only present the simulation results and the method of the γ-ray generation can be seen from the methods. More than 9×10¹³ γ-ray photons with a divergence of 3 degrees (figure 2a) are generated in each pulse. After propagating a distance (d1=70 μm) to the collision area, the pulse duration is about 21 fs and the focal spot is about 25 μm, the brilliance of the γ-ray pulse is about $1.5 \times 10^{25}$ photons·s⁻¹·mm⁻²·mrad⁻²·0.1%BW (at 0.5 MeV). The head on colliding beams luminosity[23] at different locations off the tube target end can be seen from figure 2b. In two-photon interaction (for more detail see methods), the direction of the pair is determined by the center of mass direction. Hence, the γ-ray pulse of wide-bandwidth and better collimated at higher energy (figure 2a) could be a perfect choice to create collimated pairs because most of the electron-positron pairs are generated from the interaction between two photons of different energies.

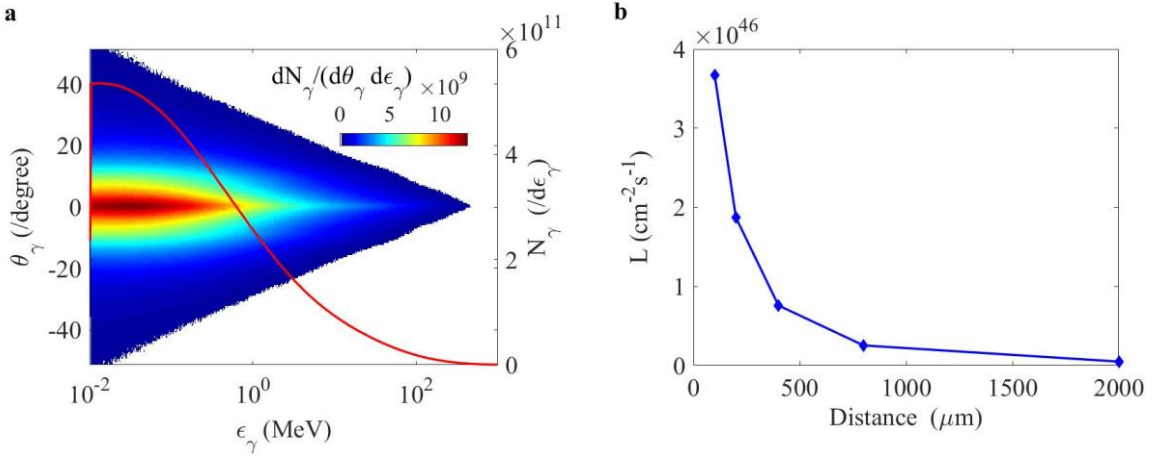

**Figure 2 | Results of gamma-ray beam. a**: The spectral-angular distribution of the gamma-ray pulse, the red line represents the energy spectrum of the gamma-ray pulse. **b**: The head on colliding beams luminosity vs. the distance d1 (figure 1).

In modelling all the possible collisions between two photon beams of a huge photon number $\sim 10^{14}$, more than $10^{28}$ interactions are needed. Such computation charges too large to be handled. The method of macro-particle, which is widely used in plasma simulation[24] for more than 50 year, can greatly reduce the computational requirement by merging a number of particles into one macro-particle. Here, we classify the photons with similar energies and momenta into a macro-photon through dividing the spectral-angular distribution (figure 2a) into segments by steps of $d\gamma = \left[\left(\log_{10}\left(\varepsilon_\gamma(\text{keV})\right)\right) * 100\right]$ and $d\theta_\gamma = 0.1$ degree. The weight value of the macro-photon is the photon number classified into it, and the energy and momentum of the macro-photon can be received by averaging all the photons in the macro-photon. Then, we can handle all the possible two-photon BW process between the macro-photons in two beams. In each interaction above the threshold condition, a macro-pair can be generated, the energy and momentum of the macro-pair is obtained by calculating the dynamics of two-photon interaction (for more detail see the methods). The weight value of the macro-pair is $w_{pair} = \frac{w_{\gamma_1} w_{\gamma_2} \sigma_{\gamma_1 \gamma_2}}{S_c}$, where $S_c$ is the size of the collision area, $\sigma_{\gamma_1 \gamma_2}$ is the cross-section of BW, $w_{\gamma_1}$ and $w_{\gamma_2}$ are the weight values of the macro-photons in the two beams.

Double γ-ray pulses (figure 2) are employed in γ-γ collider, in which the collision angle is assumed to be 170° and more than $8.7 \times 10^{10}$ ($3.2 \times 10^8$) macro-pairs (pairs) can be created in one shot. Then, we can obtain the spectral-angular distribution of the positrons (almost the same as that of the electrons) in figure 3. The positrons are better collimated at higher energy (figure 3) and the divergence angle is 7 degrees. More than 5×10⁶ positrons can be collimated into 1 degree, which are much higher than the threshold of the spectrometers[17,18] placed along the direction of the laser pulses where the signal of electron-positron pairs is maximum.

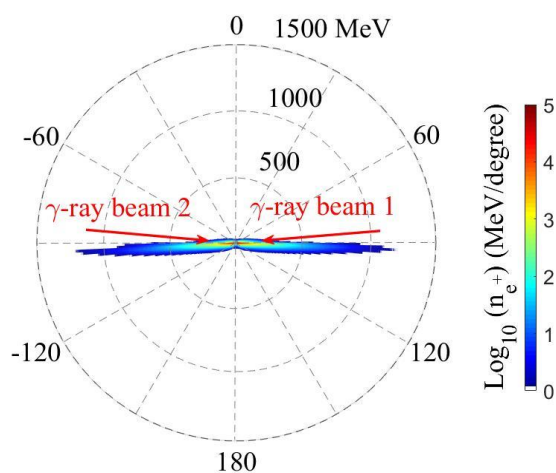

**Figure 3 | Results of the positrons.** The angular-spectrum distribution of the positrons in the case of the gamma-ray pulses driving by 10 PW laser pulses and the collision angle of $\theta_c = 170°$.

In this setup, additional electron-positron pairs can be generated from the Trident and the Bethe-Heitler processes[10,17,18] and the Triplet process[16], due to the presents of tube ions, tube electrons and the energetic electrons from the other side. All the additional positrons are considered to be the noise in this letter. The detail to estimate such positrons can be seen from the methods.

During the stage of γ-ray generation in the 400 nm gold narrow tube, the interactions between γ-ray photons and nucleus (Bethe-Heitler), γ-ray photons and electrons (Triplet), energetic electrons and nucleus (Trident), result in the positron number of 1.4×10⁷, 4.0×10⁶, 2.7×10⁶, respectively. Such positrons which move out of the tube wall due to initial transverse momenta can be deflected out of the collision area by an ultra-strong transverse electric field ($1.0\times10^{14}$ V/m) near the target out-surface (for more detail see Supplementary). Thus, the positrons generated from the narrow tube target could be safely ignored. In front of the gamma-ray pulse, there is an electron cloud generated from the other side. The density, width and charge number of the electron cloud is about $1.5\times10^{27}$ m$^{-3}$, 25 μm and <1000nC. The gamma-ray pulse collides with the electron cloud resulting in $6.4\times10^{4}$ positrons in the case of all the electrons are involved in the interaction. Hence, the BW signal is more than three orders of magnitudes higher than the noise level.

Then, we scan the production and divergence angle of the positrons for various collision angle to see the robustness of this scheme with the results presented in figure 4, from which one can see that the pair number and the divergence scale strongly with the collision angle. For a wide range of collision angle, the signals are orders of magnitude higher than the noise level. Therefore, this scheme is robustness with the experimental parameters of the up-coming laser.

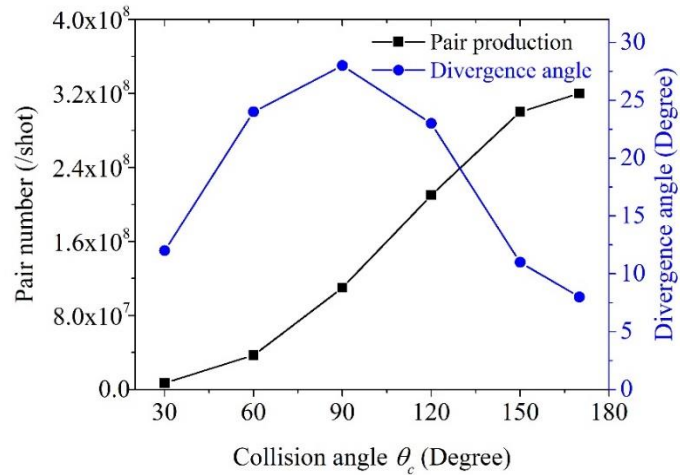

**Figure 4 | The results of the positrons (electrons).** The effect of the collision angle on the pair divergence angle (blue line) and the pair production (black line).

Conclusion, a clean channel to experimentally demonstrate the creation of electron-positron pairs in γ-γ collider is proposed by employing the γ-ray pulses generated from 10 PW laser pulses and narrow tubes. In a single shot, more than $10^8$ pairs can be created, the pairs are well collimated with a divergence of ~7 degrees, and the S/N is higher than 2000. Such scheme can realize the first observation of two photon BW process in the laboratory with the up-coming 10 PW lasers, which would significantly benefit the researches on laboratory astrophysics and quantum electrodynamics.

**Methods**

**PIC simulation setup:** We use the two-dimensional relativistic particle-in-cell code EPOCH2D[22] to model the laser plasma accelerator and the emitting of the γ-ray light source in the narrow tube target. In the PIC code, the quantum synchrotron radiation[25,26] is included, the emitting electron or positron recoils to maintain the radiation reaction force to be quantum equivalent during the photon radiation. In the laser field, the photon emission probability of the accelerated particle is governed by a parameter named optical depth $\tau_e$. The position, momentum and $\tau_e$ of the emitting particle are updated every computational time-step. More detail on the method to calculate the radiation numerically can be found in the work of Duclous et al[25]. A10-PW linearly polarized laser pulse with a transverse spatial Gaussian, $\sin^2$ temporal profile, duration of 30 fs and focal spot diameter of 2.0 μm at FWHM, irradiates into a gold narrow tube from the left side. The laser intensity $3.2 \times 10^{23}$ W/cm$^2$, corresponding to $a_0 \approx 480$, would be available in ELI[8]. The tube e$^-$/ Au$^{+69}$ density is set to 276 $n_c$ / 4 $n_c$, where $n_c$ is the critical density. The tube wall, whose thickness is 400 nm and inner diameter is 4 μm, is located 2-270 μm. The simulation box is 271 μm in the longitudinal direction (x) and 10 μm in the transverse direction (y). 5 macro-ions and 100 macro-electrons are initialized in each cell whose size is dx=dy=5nm.

**The generation of the collimated γ-ray pulse**[21]: In the interaction between the laser pulse and the narrow tube, the electrons are pulled into the narrow space by the laser transverse electric field at the beginning. Then, the electrons slip into the acceleration phase of the longitudinal electric field which accelerate the electrons longitudinally and surpass transverse momenta significantly, resulting in the collimation of the energetic electrons. Due to the transverse modulation of the electrons density, large charge separation fields (transverse electric fields) are generated near the target surface. After an acceleration distance of <100 μm, the electrons are pulled back to the tube wall by the charge separation fields which significantly enhance the γ-ray radiations near the tube wall. Hence, the γ-ray photons can be well collimated as the spread angles of the photons are determined the emitting electrons.

**Dynamics of two-photon interaction:** In the creating of electron-positron pair through the collision of two photons, the threshold condition is

$$\varepsilon_{\gamma 1}\varepsilon_{\gamma 2}(1 - \cos\theta_c) \geq 2(m_e c^2)^2, \tag{a1}$$

where $\varepsilon_{\gamma 1}$, $\varepsilon_{\gamma 2}$ are the energies of the photons, $\theta_c$ is the collision angle, $m_e$ is the electron mass at rest and $c$ is the light speed in the vacuum. The cross-section for BW process is

$$\sigma_{\gamma_1 \gamma_2} = \frac{\pi}{2} r_c^2 (1 - \beta^2) \left[ (3 - \beta^4) \ln\left(\frac{1+\beta}{1-\beta}\right) - 2\beta(2 - \beta^2) \right], \tag{a2}$$

here $r_c$ is the classical electron radius, $\beta = (1 - 1/S)^{1/2}$, $S = \varepsilon_{\gamma 1} \varepsilon_{\gamma 2} (1 - \cos\theta_c)/2(m_e c^2)^2$. After the collision, the total energy of the two photons in the frame of center-of-mass (COM) is divided equally by the electron and positron. In the COM frame, the energy of electron $\varepsilon_{ec}$ (positron $\varepsilon_{ep}$) can be expressed as

$$\varepsilon_{ec} = \varepsilon_{ep} = \frac{1}{2}\sqrt{2\varepsilon_{\gamma 1}\varepsilon_{\gamma 2}(1 - \cos\theta_c)}, \tag{a3}$$

and the electron (positron) momentum $\vec{p}_{ec}$ ($\vec{p}_{pc}$) has the following relation,

$$|p_{ec}| = |p_{pc}| = \sqrt{\frac{1}{4}\left[\left(\frac{\varepsilon_{\gamma_1}+\varepsilon_{\gamma_2}}{c}\right)^2 - (\vec{p}_{\gamma_1} + \vec{p}_{\gamma_2})^2\right] - (m_e c^2)^2} \tag{a4}$$

where $\vec{p}_{\gamma_1}$ and $\vec{p}_{\gamma_2}$ are the momenta of the gamma-ray photons in the laboratory frame. The direction of the electron (positron) momentum is considered to be isotropic in the frame of COM, but $\vec{p}_{ec} = -\vec{p}_{pc}$ must be satisfied. Through Lorentz transformation, one can get the electron (positron) energy $\varepsilon_e$ ($\varepsilon_p$) and momentum $\vec{p}_e$ ($\vec{p}_p$) in the laboratory frame:

$$\varepsilon_{e,p} = \gamma_c (\varepsilon_{ec,pc} + \vec{v}_c \cdot \vec{p}_{e,p}), \tag{a5}$$

$$\vec{p}_{e,p} = \vec{p}_{ec,pc} + \frac{\gamma_c - 1}{v_c^2}(\vec{v}_c \cdot \vec{p}_{ec,pc}) \cdot \vec{v}_c + \frac{\gamma_c \vec{v}_c \varepsilon_{ec,pc}}{c^2}, \tag{a6}$$

where $\gamma_c = 1/\sqrt{1 - (v_c/c)^2}$, the velocity of the COM $\vec{v}_c$ can be expressed as

$$\vec{v}_c = \frac{(\vec{p}_{\gamma_1} + \vec{p}_{\gamma_2})c^2}{\varepsilon_{\gamma_1} + \varepsilon_{\gamma_2}} \tag{a7}$$

**Pair productions from Trident, Bethe-Heitler, Triplet:** In the Trident process, electron-positron pairs can be generated by the colliding between the fast electrons and the Coulomb field of the atomic nucleus. The total cross-section[27] of Trident can be expressed as

$$\sigma_T = \begin{cases} 5.22 Z^2 \log^3 \left[\frac{2.3 + T_0(\text{MeV})}{3.52}\right] \mu b & (T_0 < 100 \text{MeV}) \\ \frac{28\pi Z^2 r_c^2 \alpha^2}{27} \log^3 \left[\frac{T_0(\text{MeV})}{m_e c^2}\right] \mu b & (T_0 > 100 \text{MeV}) \end{cases}, \tag{b1}$$

where $T_0$ represents the kinetic energy of the electrons, $\alpha = 1/137$ is the constant of fine-structure. The Trident pair number can be estimated by

$$N_{T\pm} = \sum_{i=1}^{N_e} \frac{\sigma_T N_z}{S_C} = \sum_{i=1}^{N_e} \sigma_T d n_z, \tag{b2}$$

where $N_e$, $N_z$ are the total number of electrons and ions involved in the interaction, $d$ is the thickness of the carbon film and $n_Z$ is the number density of the high-Z material.

In the Bethe-Heitler process, electron-positron pairs can be created from the interaction between γ-ray photons and the Coulomb field of the atomic nucleus. The cross-section[27,28] of Bethe-Heitler is

$$\sigma_{BH} = \frac{\sigma_T}{\frac{\alpha}{\pi}\left[\log\left(\frac{\varepsilon_0}{m_e c^2}\right)\log\left(\frac{\varepsilon_0}{2.137 m_e c^2 Z^{-1/3}}\right) + \frac{1}{3}\log^2(2.137 Z^{-1/3})\right]}, \quad (b3)$$

where Z is the atomic number, and $\varepsilon_0 \approx T_0$ is the energy of the photons. The Bethe-Heitler pair production can be calculated by

$$N_{BH\pm} = \sum_{i=1}^{N_\gamma} \sigma_{BH} dn_Z, \quad (b4)$$

here $N_\gamma$ is the number of the γ-ray photons.

In the mechanism of Triplet production[16], the cross-section becomes much more complicated. To cover the photon energy of whole range from the threshold, Haug[29] gave the following analytic expressions of the cross-section from threshold,

$$\frac{\sigma_{\gamma e}}{\alpha r_c^2} = \begin{cases} [-76 + 20.4k - 10.9(k-4)^2 - 3.6(k-4)^3 + 7.4(k-4)^4] \cdot 10^{-3}(k-4)^2, (k \leq 4.6) \\ 0.582814 - 0.29842k + 0.04354k^2 - 0.0012977k^3, (4.6 \leq k \leq 6.0) \\ (3.1247 - 1.3394k + 0.14612k^2)/(1 + 0.4648k + 0.016683k^2), (6.0 \leq k \leq 18) \\ \frac{28}{9}\ln(2k) - \frac{218}{27} + \frac{1}{k}\left[27.9 - 11\ln(2k) + 3.863\ln^2(2k) - \frac{4}{3}\ln^3(2k)\right], (k \geq 14) \end{cases}, \quad (b5)$$

where $k = \frac{\varepsilon_0}{m_e c^2}$. The pair production from Triplet process is

$$N_{\gamma e\pm} = \sum_{i=1}^{N_\gamma} \sigma_{\gamma e}(dn_Z + N_{e2}), \quad (b6)$$

here $N_{e2}$ is the electron number contributed to the interaction from the other electron bunch.

**Acknowledgements**

The work has been supported by the National Basic Research Program of China (Grant No.2013CBA01502), NSFC (Grant Nos. 11475010, 11575011, 11535001), National Grand Instrument Project (2012YQ030142), the Projects (2016M600007, 2017T100009) funded by China Postdoctoral Science Foundation. The PIC code Epoch was in part funded by the UK EPSRC grants EP/G054950/1. Simulations were supported by High-performance Computing Platform of Peking University. The authors also thank Professor S. J. Rose and Professor Z. Najmudin from Imperial College London, and Doctor W. R. Chou from Fermilab for their helpful comments. Y. S. H. thanks the support by the CAS Center for Excellence in Particle Physics (CCEPP)

**Author contributions**

J. Q. Y. designed this research, carried out the simulation and calculations, and prepared the manuscript. H. Y. L. designed this research prepared the manuscript. T. T. worked on the calculations. X. Q. Y. leaded this project, designed this research and prepared the manuscript. All the authors contributed in finalizing the manuscript.

**Competing financial interests**

The authors declare no competing financial interests.

# Supplementary

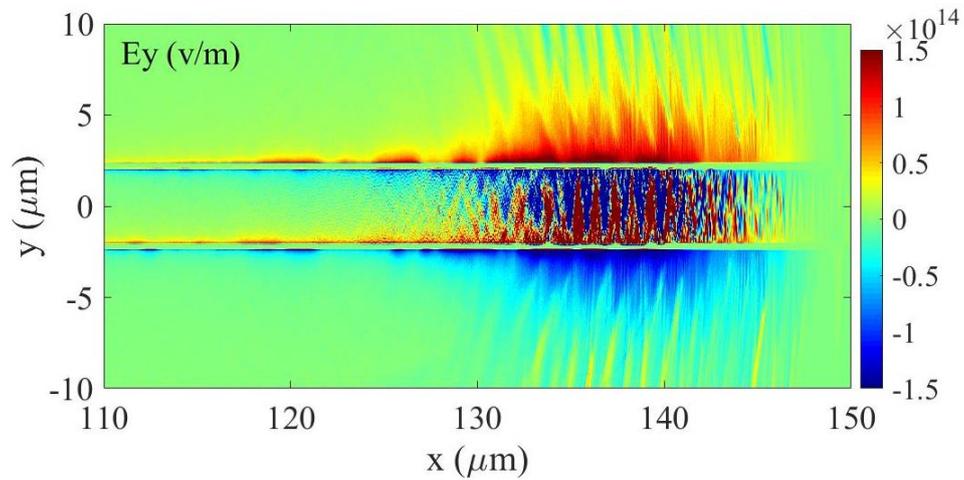

**Figure 1 | Transverse field distribution at 150 fs into the simulation**: When the laser pulse transports in the narrow tube target. Electrons are accelerated transversely. Such electrons penetrate through the target and an ultra-strong charge separation field is formed near the surface of the target. The positrons, which move into the charge separation field due to initial transverse momenta, will obtain a large transverse momentum.